\documentclass[12pt]{article}

\usepackage{subfigure}

\usepackage{latexsym}
\usepackage{graphics}
\usepackage{amsmath}
\usepackage{xspace}
\usepackage{amssymb}
\usepackage{psfrag}
\usepackage{epsfig}
\usepackage{amsthm}
\usepackage{pst-all}

\usepackage{color}

\definecolor{Red}{rgb}{1,0,0}
\definecolor{Blue}{rgb}{0,0,1}
\definecolor{Olive}{rgb}{0.41,0.55,0.13}
\definecolor{Yarok}{rgb}{0,0.5,0}
\definecolor{Green}{rgb}{0,1,0}
\definecolor{MGreen}{rgb}{0,0.8,0}
\definecolor{DGreen}{rgb}{0,0.55,0}
\definecolor{Yellow}{rgb}{1,1,0}
\definecolor{Cyan}{rgb}{0,1,1}
\definecolor{Magenta}{rgb}{1,0,1}
\definecolor{Orange}{rgb}{1,.5,0}
\definecolor{Violet}{rgb}{.5,0,.5}
\definecolor{Purple}{rgb}{.75,0,.25}
\definecolor{Brown}{rgb}{.75,.5,.25}
\definecolor{Grey}{rgb}{.5,.5,.5}

\newcommand{\E}[1]{\mathbb{E}\!\left[#1\right]}
\newcommand{\R}{\mathbb{R}}

\newcommand{\G}{\mathbb{G}}

\newcommand{\ignore}[1]{\relax}

\newcommand{\ER}{Erd{\"o}s-R\'{e}nyi }

\definecolor{Red}{rgb}{1,0,0}
\definecolor{Blue}{rgb}{0,0,1}
\definecolor{Olive}{rgb}{0.41,0.55,0.13}
\definecolor{Green}{rgb}{0,1,0}
\definecolor{MGreen}{rgb}{0,0.8,0}
\definecolor{DGreen}{rgb}{0,0.55,0}
\definecolor{Yellow}{rgb}{1,1,0}
\definecolor{Cyan}{rgb}{0,1,1}
\definecolor{Magenta}{rgb}{1,0,1}
\definecolor{Orange}{rgb}{1,.5,0}
\definecolor{Violet}{rgb}{.5,0,.5}
\definecolor{Purple}{rgb}{.75,0,.25}
\definecolor{Brown}{rgb}{.75,.5,.25}
\definecolor{Grey}{rgb}{.5,.5,.5}
\definecolor{Pink}{rgb}{1,0,1}
\definecolor{DBrown}{rgb}{.5,.34,.16}
\definecolor{Black}{rgb}{0,0,0}

\usepackage{float}

\author{
{\sf David Gamarnik
\thanks{Operations research Center and Sloan School of Management, MIT. Email: gamarnik@mit.edu} 
\thanks{Support from the NSF grant DMS-2015517 is gratefully acknowledged.}
}
}

\begin{document}

\title{The Overlap Gap Property: a Topological Barrier to  Optimizing over  Random Structures}
\date{\today}

\maketitle

\begin{abstract}
The problem of optimizing over random structures emerges in many areas of science and engineering, ranging from statistical physics to machine learning
and artificial intelligence. For many such structures finding optimal solutions by means of fast algorithms 
is not known and often is  believed not possible.
At the same time the formal hardness of these problems in form of say complexity-theoretic $NP$-hardness is lacking. 

In this introductory article a new approach for algorithmic
intractability in random structures is described, which is based on the topological disconnectivity property of the set of  
pair-wise distances of near optimal solutions, called the Overlap Gap Property. 
The article demonstrates how this property a) emerges in most models known to exhibit an apparent 
algorithmic hardness b) is consistent with the hardness/tractability
phase transition for many models analyzed to the day, 
and importantly c) allows to mathematically rigorously rule out  large classes 
of algorithms as potential contenders, in particular the 
algorithms  exhibiting the input stability (insensitivity). 
\end{abstract}



Optimization problems involving uncertainty emerge in many areas of science and engineering, including statistics, machine learning and artificial
intelligence, computer science, physics, biology, management science, economics and social sciences. The exact nature and the sources of uncertainty 
vary from field to field. The 
modern paradigm of Big Data brought forward 
  optimization problems involving in particular many dimensions, thus   creating a new
meta-field of ''high dimensional
probability'' and ''high-dimensional statistics''~\cite{vershynin2018high},\cite{buhlmann2011statistics},\cite{foucart2013mathematical}.
Two recent special semester programs at the Simons Institute for the Theory of Computing were devoted to this burgeoning 
topic~\footnote{https://simons.berkeley.edu/programs/si2021,
https://simons.berkeley.edu/programs/hd20}, among
a multitude of other conferences and workshops. This paper is based on several lectures given by the author
in a reading group during the first of these events.
 While many of the optimization problems involving randomness 
can be solved to optimality by fast, and in fact sometimes border-line trivial algorithms, other problems have resisted decades of attempts and slowly it has been
accepted  that these problems are likely non-amenable to fast algorithms, with polynomial time algorithms broadly considered to be the  
gold standards for 
what constitutes fast algorithms.  
The debate surrounding the actual algorithmic hardness of these problems though
is by no means settled, unlike its ''worst-case'' algorithmic complexity-theoretic counterparts,
expressed in the form  of the widely believed  $P\ne NP$ conjecture. 

What is the ''right'' algorithmic complexity theory explaining the persistent failure
to find tractable algorithms for these problems? We discuss in this paper some existing theories and explain their shortcomings 
in light what we know now about the state of the art algorithms. 
We then propose an approach, largely inspired by the field of statistical physics, which offers a new topological/geometric
theory of algorithmic hardness that is 
based on the disconnectivity of the overlaps of near optimal solutions, dubbed the ''Overlap Gap Property'' (OGP). The property has been rigorously verified for many concrete models, and, importantly 
can be used to mathematically rule out large classes of algorithms as potential contenders,
specifically algorithms exhibiting a form of input stability. This includes both classical and quantum algorithms,
the latter class including algorithms called QAOA 
(Quantum Approximate Optimization Algorithm),~\cite{farhi2020quantumRandom},\cite{farhi2020quantumWorstCase} 
which gained recently a lot of attention
as the most realistic algorithm to be implementable on quantum computers~\cite{farhi2014quantum}.
A widely studied random Number Partitioning problem will be used to illustrate both the OGP 
 and the mechanics by which this property
manifests itself as a barrier for  tractable algorithms. 

\section*{The Largest Clique of a Random Graph: the Most ''Embarrassing'' Open Problem in Random Structures}
Imagine a club with $N$ members in which about 50\% of the $N(N-1)/2$ member pairs know each other personally, and the remaining
50\% of the members do not. You want to find a largest clique
in this club, namely the largest group  of members out of the $N$ members
who all know each other. What is the typical size $c^*$ of such a clique? How easy is it to find one?
This question can be modeled as the problem of finding a largest fully connected sub-graph (formally called a ''clique'') in a random
\ER graph $\G(N,1/2)$, which is a graph on $N$ nodes where every pair of nodes is connected with probability $1/2$, independently for all pairs. 
The first question regarding the largest clique size $c^*$ is a textbook example of the so-called probabilistic method, 
a simple application of which tells us that $c^*$ is likely to be near $2\log_2 N$ with high degree of certainty as $N$ 
gets large~\cite{alon2004probabilistic}. 
A totally different matter is the problem of actually finding such a clique and this is where the embarrassment begins. Richard Karp, one of the founding fathers
of the algorithmic complexity theory, observed in his 1976 paper~\cite{karp1976probabilistic} 
that a very simple algorithm, both in terms of the analysis and the implementation,
finds a clique of roughly half-optimum size, namely with about $\log_2 N$ members, and challenged  to do better. The problem 
is still open and this is embarrassing for two reasons: (a) The best known algorithm, namely the one above,
is also an extremely naive one.  So it appears that the significant progress that the algorithmic community has 
achieved 
over the past decades in constructing ever more clever and powerful algorithms, is totally helpless in improving upon this extremely simple and naive
algorithm; (b) We don't have a workable
theory of algorithmic hardness which rigorously explains why finding cliques of size half-optimum is easy, 
and improving on it does not seem  possible within the
realm of polynomial time algorithms. The classical $P\ne NP$ paradigm and its variants is of no help here, more on this below. 
An approach based on the solution space landscape
can be used to rule out Markov Chain Monte Carlo (MCMC) algorithms for finding cliques larger than half-optimum 
by showing that the associated chain mixes at scale
larger than polynomial time, and thus ruling out the MCMC algorithm~\cite{jerrum1992large}. 
The method however does not appear extendable
to other types algorithms which are not based on MCMC method.

The largest clique problem turns out to be one of very many other problems 
exhibiting a similar phenomena: using non-constructive analysis method
one shows that the optimal value of some optimization problem involving 
randomness is some  value $c^*$,  the best known polynomial time 
algorithm achieves value $c_{\rm ALG}$, and there is a significant gap 
between the two: $c_{\rm ALG}<c^*$. A partial list (but one growing seemingly
at the speed of the Moore's Law) is the following: randomly generated 
constraint satisfaction problem (aka the random  K-SAT problem), 
largest  independent set in a random graph, proper coloring of a random graph, finding a ground state of a spin 
glass model in statistical mechanics, discovering communities in 
a network (the so-called community detection problem), group testing,
statistical learning of a  mixture of Gaussian distributions, sparse linear regression problem, 
sparse covariance estimation, graph matching problem,
spiked tensor problem, the Number Partitioning problem, and many other problems.

The Number Partitioning problem is motivated in particular by the statistical 
problem of designing a randomized control study with two groups possessing
roughly ''equal'' attributes. It is also a special case of the bin-packing problem, and also has been
widely studied in the statistical physics literature. The Number Partitioning problem will serve as 
one of the  running
example in this article, illustrating our main ideas, so we introduce it now. Given $N$ items with 
weights $X_1,\ldots,X_N$, the  goal is to split it into
two groups so that the difference of total weights in two groups is the smallest possible. 
Namely, the goal is finding a subset $I\subset 1,\ldots,N$ such 
that $|\sum_{i\in I}X_i-\sum_{i\notin I}X_i|$ is as small as possible. 
An NP-hard problem in the worst case~\cite{garey1979computers}, 
it is more
tractable in  presence of randomness. Suppose, the weights $X_i$ are generated 
independently according to the standard Gaussian distribution $\mathcal{N}(0,1)$. 
A rather straightforward application of the same probabilistic method 
shows that the optimum value is typically $\sqrt{N}2^{-N}$ for large $N$. 
Namely, the value of $c^*$
in our context is $\sqrt{N}2^{-N}$.
An algorithm proposed by Karmarkar and Karp in 1982~\cite{karmarkar1982differencing}
achieves the value of order $c_{\rm ALG}=\sqrt{N}e^{-c\log_2^2 N}$, 
with value $c$ predicted (though not proven) to be $1/(2\log 2)\approx 0.721..~$~\cite{boettcher2008analysis}. 
The multiplicative
gap between the two is thus very significant (exponential): $c_{\rm ALG}/c^*=2^{N-O(\log_2^2 N)}$. 
While multidimensional extensions of this algorithm have been considered 
recently~\cite{turner2020balancing},\cite{harshaw2019balancing}, 
no improvement of this result is known  to the date,
and no algorithmic hardness result is known either.

\section*{In Search of the ''Right'' Algorithmic Complexity Theory}
We now describe some existing approaches for understanding the algorithmic complexity and discuss their shortcomings in explaining
the existential/algorithmic gap exhibited by the multitude of problems described above. Starting with the  classical algorithmic complexity theory
based on the algorithmic complexity classes such as $P$, $NP$, etc. this theory is of no help, since these complexity classes 
are based on the worst-case assumptions
on the problem description. For example, assuming the widely believed conjecture that $P\ne NP$, finding a largest clique in a graph 
with $N$ nodes within a multiplicative factor $N^{1-\delta}$ is not possible by polynomial time 
algorithms~\cite{Hastad} for any constant $\delta\in (0,1)$ if $P\ne NP$. 
This  is a statement, however, about the algorithmic problem
of finding large cliques in \emph{all} graphs, 
and it is in sharp contrast with the factor $1/2$ achievable by polynomial time algorithms 
for \emph{random} graphs $\G(N,1/2)$, according to 
the discussion above.

A significant part of the algorithmic complexity theory   
does in fact deal with problems with random input and here 
an impressive random-case to worst-case 
reductions are achievable for some of the problems. 
These problems enjoy wide range applications in cryptography, where the 
average case hardness property
is paramount for designing secure cryptographic protocols.
 For example, if there exists a polynomial time 
algorithm for computing the permanent of a matrix with i.i.d. 
random inputs, then, rather surprisingly,  one can use it to design a polynomial time algorithm 
for \emph{all} matrices~\cite{lipton1991new}.
The latter problem is known to be in the $\#P$ complexity class which subsumes $NP$. 
A similar reduction exists~\cite{gamarnik2018computing} for the problem of computing the 
partition function of a Sherrington-Kirkpatrick model described below, 
thus implying
that computing partition functions for spin glass models is not possible by polynomial time algorithms unless $P=\#P$. 
Another problem admitting average-case to worst-case reduction
is the problem of finding a shortest vector in a lattice~\cite{ajtai1996generating}. 
The random to worst case types of reduction described above would be ideal for 
our setting, as they would provide the most compelling 
evidence of hardness of these problems. For example, it would be ideal to 
show that finding a clique of size at least say $(3/2)\log_2 N$ in $\G(N,1/2)$ 
implies a polynomial time algorithm with the same approximation 
guarantee for all graphs. Unfortunately, such a result appears out of reach for 
the existing proof techniques. 

Closer to the approach described in this paper, 
and the one which in fact has largely inspired the present line of work, 
is a range of theories based on the solution
space geometry of the underlying optimization problem. 
This approach emerged in the statistical physics literature, specifically the study of spin glasses,
and more recently found its way to questions above and beyond statistical physics, in particular questions
arising in the context of studying neural networks as~\cite{choromanska2015loss}.
The general philosophy of this take on the algorithmic complexity 
is that when the problem appears to be algorithmically hard, this somehow should be reflected
in the non-trivial geometry of optimal or near optimal solutions. 
One of the earliest such approaches was formulated for the decision
problems, such as random constraint satisfaction problems (aka random K-SAT problem).
It links the algorithmic hardness with the proximity to the satisfiability phase transition threshold~\cite{fu1986application},
\cite{kirkpatrick1994critical}, and the order (first vs second) of 
the associated phase transition~\cite{monasson1999determining}. 
To elaborate, we need to introduce the
random K-SAT problem first. It is a Boolean constraint satisfaction 
problem involving $N$ variables $x_1,\ldots,x_N$ defined as a conjunction $\Phi$ of $M$
clauses $C_1\wedge C_2\wedge \cdots \wedge C_M$, where each 
clause $C_j$ is a disjunction of exactly $K$ variables from $x_1,\ldots,x_N$ or their negation.
Thus, each $C_j$ is of the form $x_{j,1}\vee \neg x_{j,2}\vee\cdots \vee \neg x_{j,K}$. 
An example of such a formula with $N=10$, $M=4$ and $K=3$ is say
$(x_2\vee \neg x_5\vee x_6)\wedge (\neg x_1\vee \neg x_2\vee x_9)\wedge (\neg x_6\vee  x_8\vee \neg x_{10})\wedge (\neg x_3\vee  x_2\vee x_7)$. 
A formula is called satisfiable if there exists an assignment of Boolean variables
$x_i, 1\le i\le N$ to values $0$ and $1$ such that the value of the formula is $1$. 

A random instance of a K-SAT problem is obtained by selecting variables 
into each clause $C_j$ uniformly at random, independently for all $j$
and applying the negation operation $\neg$ with probability $1/2$ 
independently for all variables. Random K-SAT problem is viewed
by statistical physicists as a problem exhibiting the so-called frustration property, 
which is of great interest to physics of spin glasses,
and hence it has enjoyed a great deal of attention in the statistical physics 
literature~\cite{MezardParisiVirasoro},\cite{MezardMontanariBook}.

As it turns out, for each $K$ there is a conjectured 
critical threshold $\alpha_{\rm SAT}=M/N$ for the
satisfiability property, which was rigorously proven~\cite{ding2015proof} for large enough $K$. 
In particular, for every $\alpha<\alpha_{\rm SAT}$,  the formula admits a satisfying 
assignment when $M/N<\alpha$, and for every $\alpha>\alpha_{\rm SAT}$, 
the formula does not admit a satisfying assignment when $M/N>\alpha$, both 
with overwhelming probability as $N\to\infty$. 
The sharpness of this transition was established rigorously earlier
by general methods in~\cite{Friedgut}. 
The algorithmic counterparts, however, came short of achieving
the value $\alpha_{\rm SAT}$ for every $K\ge 3$ (more on this below).
Even before the results in~\cite{ding2015proof} and \cite{Friedgut}, it was conjectured 
in the physics literature  that perhaps the existence
of the sharp phase transition property itself 
is the culprit for the algorithmic hardness~\cite{kirkpatrick1994critical}, 
\cite{monasson1999determining}. This was argued by studying the heuristic running times of finding the satisfying
solutions or demonstrating non-existence of ones for small values of $K$, 
and observing that the running time explodes near $\alpha_{\rm SAT}$, and subsides
when $M/N$ is far from $\alpha_{\rm SAT}$ on either side. The values of
$\alpha_{\rm SAT}$ are known thanks to the powerful machinery of the replica symmetry methods developed in a seminal works of 
Parisi~\cite{parisi1980sequence},\cite{MezardParisiVirasoro}. These methods give very precise
predictions for the values of $\alpha_{\rm SAT}$ for every $K$.
For example $\alpha_{\rm SAT}$ is approximately
$4.26$ when $K=3$. An in depth analysis of the exponent of the running times was reported in~\cite{kirkpatrick1994critical}. 

The theory postulating that the algorithmic
hardness of the K-SAT problem is linked with the satisfiability phase transition however appears inconsistent with the later 
rigorous discoveries  obtained specifically for large values of $K$. 
In particular, while $\alpha_{\rm SAT}$ is known to be approximately $2^K\log 2$ for large
values of $K$, all of the
known algorithms stall long before that, 
and specifically at $\alpha_{\rm ALG}=(2^K/K)\log K$~\cite{coja2010better}. Furthermore, there is an evidence
that breaking this barrier might be extremely challenging. This was argued by proving 
that the model undergoes the so-called \emph{clustering}
phase transition near $\alpha_{\rm ALG}$~\cite{achlioptas2006solution},\cite{mezard2005clustering} 
(more on this below), but and also by ruling out various families of algorithms. These algorithms include
sequential local algorithms and algorithms based on low-degree polynomials (using the Overlap Gap Property 
discussed in the next section)~\cite{gamarnik2017performance},\cite{bresler2021algorithmic}, 
the Survey Propagation algorithm~\cite{hetterich2016analysing}, 
and a variant of a random search algorithm called WalkSAT~\cite{coja2017walksat}. 
The Survey Propagation algorithm is a highly effective heuristics for finding
satisfying assignments in random K-SAT  and many other similar problems.
It is particularly effective for low values of $K$ and
beats many other existing approaches in terms of the running times
and sizes of instances it is capable 
to handle~\cite{mezard2002analytic} (see also~\cite{gomes2002satisfied} for the perspective on the approach).
It is entirely possible that this algorithm and even its more elementary
version, the Belief Propagation guided decimation algorithm solves the satisfiability problem for small
values of $K$. An evidence of this can be found in~\cite{ricci2009cavity}, and rigorous analysis
of this algorithm was conducted by Coja-Oghlan in~\cite{coja2011belief}, which shows it to be effective
when $\alpha\le 2^K/K$.
However, the analysis by Hetterich in~\cite{hetterich2016analysing} rules out the Survey Propagation 
algorithm for sufficiently large values of $K$, beyond the $\alpha_{\rm ALG}$ threshold, which 
we recall is $(2^K/K)\log K$.
Additionally,  the theory 
of algorithmic hardness based on the existence of the satisfiability 
type phase transition does not appear to explain the algorithmic hardness
associated with  \emph{optimization} problems,
such as the largest clique problem in $\G(N,1/2)$.

\begin{figure}
\begin{center}
\scalebox{.6}{\includegraphics{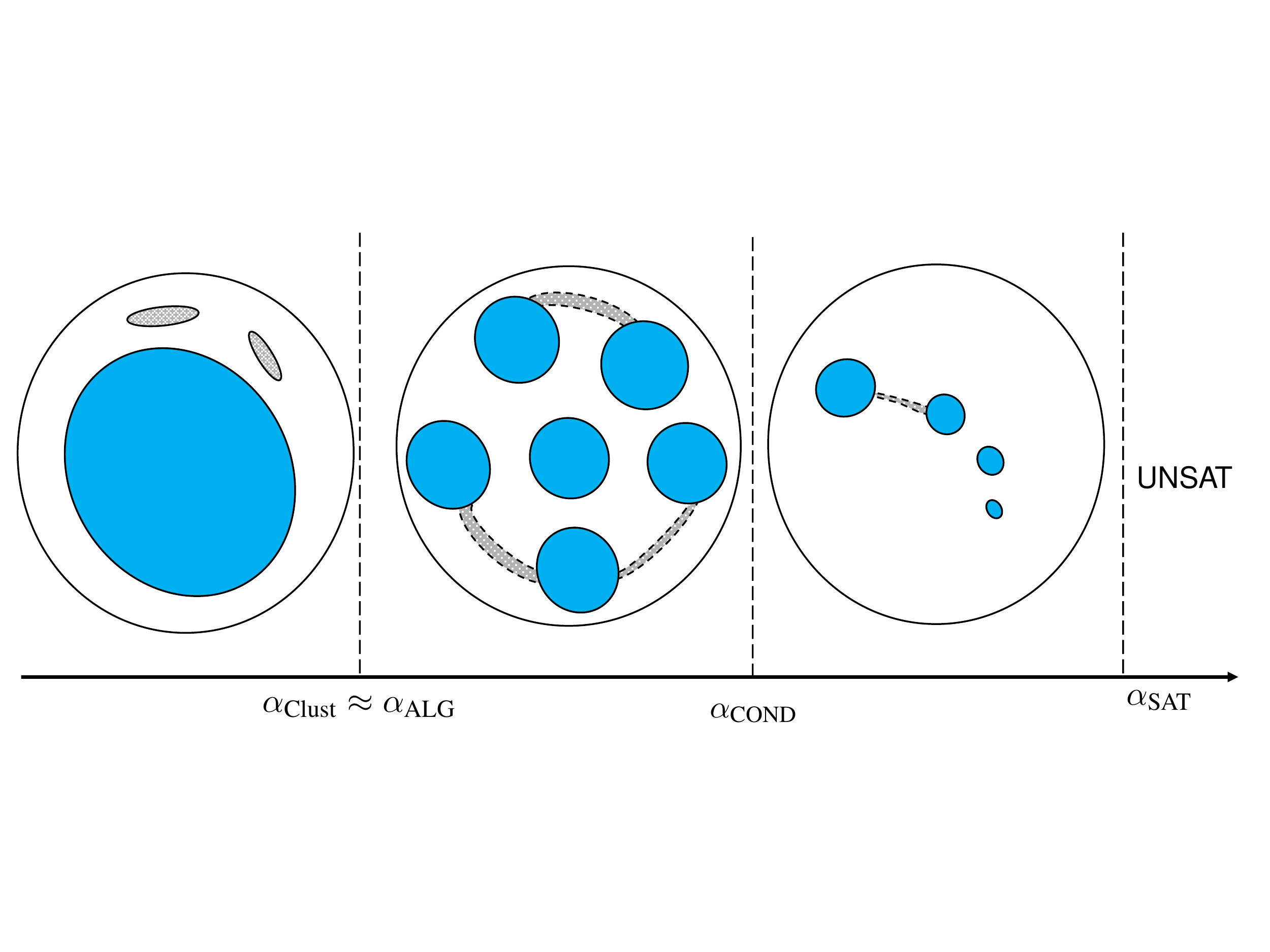}}
\end{center}
\caption{Clustering phase transition. Clustered 
sets are represented by blue colors and the exception sets are represented  by the grey colored tunnels}
\label{fig:clust-cond}
\end{figure}

The clustering property was rigorously established~\cite{achlioptas2006solution},\cite{mezard2005clustering} 
for  random K-SAT
for values of $\alpha$ above $\alpha_{\rm Clust}$, which is known to be close 
to $\alpha_{\rm ALG}$ for large  $K$. We specifically refer to this as \emph{weak clustering}
property in order to distinguish it from the \emph{strong} clustering property, both of which we now 
define. (While the distinction between weak and strong clustering was discussed in many prior works,
the choice of the terminology is entirely by the author). The model exhibits the weak clustering property 
if there exists a subset $\Sigma$ of satisfying assignments which contains 
all but exponentially small in $N$ fraction of  the satisfying assignments, 
and which can be partitioned into subsets (clusters)
separated by order $O(N)$ distances, such that within each cluster one can move between any two assignments 
by changing at most constantly many $O(1)$ bits. In other words, by declaring two assignments $\sigma$
and $\tau$ connected if $\tau$ can be obtained from $\sigma$ by flipping the values of at most $O(1)$ variables, the 
set $\Sigma$ consists of several disconnected components separated by linear in $N$ distances.
This is illustrated on Figure~\ref{fig:clust-cond} where blue regions represent clusters in $\Sigma$.
The grey ''tunnel'' like sets depicted on the figure 
are part of the  complement (exception) subset $\Sigma^c$ of the set of satisfying assignments,
which may potentially  connect (tunnel between) the clusters.
Furthermore, the clusters are separated by exponentially large in $N$ cost barriers, meaning that any path 
connecting two solutions from different clusters, (which then necessarily  contains assignments violating
at least some of the constraints), at some point contains in it  assignments which in fact violate order $O(N)$ constraint.
For this reason the clustering property is sometimes referred to as ''energetic'' barrier.

As mentioned earlier, for large values of $K$ the onset of weak clustering occurs near $2^K\log 2/K$,
and in fact the suggestion that this value indeed corresponds to the algorithmic threshold $\alpha_{\rm ALG}$
was partially motivated by this discovery of weak clustering property around this value. 
See also~\cite{ricci2010being} for the issues connecting  
the clustering property with the algorithmic hardness. Below $\alpha_{\rm ALG}$ the bulk of the set of 
satisfying assignments  constitutes one connected subset.

On the other hand, the \emph{strong} clustering property is the property that \emph{all} satisfying
assignments can be partitioned into clusters like the ones  above, however, with no exceptions. Namely $\Sigma$ is the 
set of \emph{all} satisfying assignments. This can be visualized from Figure~\ref{fig:clust-cond}
where the grey ''tunnels'' are removed.
It turns out that for large values of $K$, the model does 
exhibit the strong clustering property as well, but the known lower bounds for it are of the order
$2^K$ as opposed to $2^K/K$, for large $K$. While not proven formally, it is expected though that 
the onset of the strong clustering property indeed occurs at values order $2^K$ as opposed to $2^K/K$ for large $K$.
Namely the  weak and strong clustering properties appear at  different in $K$ scales.

The existence of the strong clustering property
is established as an implication of the Overlap Gap Property, (which is the subject of this paper) and this 
was the approach used in~\cite{achlioptas2006solution},\cite{mezard2005clustering}. In contrast the
weak clustering property is established by using the so-called \emph{planting} solution
which amounts to consider the solution space geometry from the perspective of a uniformly at  random selected
satisfying assignment.

The necessity of distinguishing between the two modes of clustering described above 
is driven by algorithmic implications.
Solutions generated by most algorithms typically are not generated uniformly at random and thus in principle can
easily fall into the exception set $\Sigma^c$. Thus the obstruction arguments based somehow on linear separations
between the clusters and exponentially large energetic barriers might simply be of no algorithmic relevance.
This is elaborated further in later section devoted to the relationship between clustering and the 
Overlap Gap Property using specifically the binary perceptron model example.
In contrast, the strong clustering property implies that \emph{every} solution produced by the algorithm belongs to one
of the clusters. Thus if two implementations of the algorithm produce solutions in different clusters, 
then informally the algorithm is capable of  ''jumping'' over large distances between the clusters. In fact this intuition 
is formalized using the Overlap Gap Property notion by means of  extending the strong clustering
property to an ensemble of correlated instances (which we call ensemble Overlap Gap Property) pitted against
the stability (input insensitivity) of algorithms. This is discussed in depth in the next section.

As the notion of ''exception'' size in the definition of weak clustering property hinges on the 
counting measure associated with the space of satisfying assignments (and the uniform measure was assumed in the
discussion above) a potential improvement might stem from changing this counting measures and introducing solution
specific weights. Interestingly, for small values of $K$ this can be done effectively delaying the onset of weak
clustering, as was shown by Budzynski et al in~\cite{budzynski2019biased}. Unfortunately, though, for large $K$
the gain is only in second order terms, and up to those orders, the threshold for the onset of weak clustering
based on biased counting measures is still $2^K\log 2/K$ as shown by Budzynski and Semerjian in~\cite{budzynski2020biased}.
This arguably, provides an even more robust  evidence that this value marks a phase transition point of fundamental nature.

More recently  algorithmic barriers were suggested to be linked with refined properties of the 
clustering phase transition, specifically
the so-called condensation phase transition~\cite{krzakala2007gibbs} 
and the emergence of so-called
\emph{frozen} variables. Algorithmic implications of both are discussed in~\cite{marino2016backtracking}. 
Using again a powerful machinery of the replica symmetry breaking, 
providing non-rigorous, though highly believable
predictions, one identifies another phase transition $\alpha_{\rm COND}$ satisfying 
$\alpha_{\rm ALG}<\alpha_{\rm COND}<\alpha_{\rm SAT}$ \emph{for large $K$}. 
As the value $M/N$ passes through $\alpha_{\rm COND}$  
the number of solution clusters covering the majority of satisfying assignment
in the random K-SAT problem drops dramatically from the exponentially many  to only constantly many ones, 
with the largest cluster containing a non-trivial
fraction of all of the assignments~\cite{krzakala2007gibbs}, see Figure~\ref{fig:clust-cond}.
At this stage a long-range dependence between the spin magnetization (appropriately defined) emerges and furthermore
at this stage the random overlaps (inner products) of pairs 
of assignments generated according to the associated Gibbs measure have a non-trivial
limiting distribution, described by the so-called Parisi measure. 

Even before the value  $M/N$ exceeds the value $\alpha_{\rm COND}$ for large $K$,
each cluster contains a non-trivial fraction  of the 
frozen variables, which are variables always taking the same 
values within a given cluster of satisfying assignment.
The authors in \cite{marino2016backtracking} conjecture that 
the emergence of frozen variables is the primary culprit of the algorithmic hardness and 
construct a variant of the Survey Propagation algorithm called Backtracking Survey Propagation algorithm which is conjectured
to be effective all the way up to  $M/N<\alpha_{\rm COND}$. 
Numerical simulations
demonstrate an excellent performance of this algorithm on randomly generated instances of K-SAT for small values of $K$.

The conjecture linking the emergence of frozen variables with 
algorithmic hardness stumbles however upon similar challenges as the theory based
on the existence of the phase transition. A rigorous evidence has been established that when $K$ is large, 
$\alpha_{\rm COND}$ is of the order $2^K\log 2-C$
for some (explicitly computable) value $C$ which does not depend on $K$. 
In particular $\alpha_{\rm COND}$ is substantially above
the value $\alpha_{\rm ALG}$ at which all known algorithms fail and 
furthermore classes of algorithms, including the standard Survey Propagation algorithm,
are ruled out as discussed above. 
Of course the non-existence of evidence is not necessarily the  evidence of non-existence, and it is very much
possible that in future successful polynomial time algorithms will be constructed in the 
regime $\alpha_{\rm ALG}<M/N<\alpha_{\rm COND}$. 
But there is another limitation  of the theory of algorithmic hardness based on the emergence of frozen variables: 
just like the notion of satisfiability phase transition,  
the notion of frozen variables does not apply to problems of optimization type, 
such as the largest clique problem, or  optimization problem appearing in the context of spin glass models, for example
the problem of finding a ground state of a  spherical $p$-spin glass model. 
In this model spins take values in a continuous range and thus
the notion of frozenness, which is exclusively discrete, is lost.

\section*{Topological Complexity Barriers. The Overlap Gap Property}
We now introduce a new approach for predicting and proving  
 algorithmic complexity for solving random constraint satisfaction problems and optimization problems
on  random structures, such as the ones introduced in earlier sections. The approach bears a lot of similarity with the predictions based
on the clustering property described in the earlier section, but has important and nuanced differences, 
which  allows us to rule out
large classes of algorithms, something that did not seem  possible before.
An  important special case  of such algorithms are stable algorithms, namely algorithms exhibiting low sensitivity to the data input. 
A special yet very broad and powerful class of such algorithms are algorithms based on low-degree polynomials.
A fairly recent stream of research \cite{hopkins2017efficient},\cite{hopkins2017power},\cite{hopkins2018statistical}
puts forward an evidence that in fact algorithms based on low-degree polynomials might be the most
powerful class of polynomial time algorithms for optimization problem on random structures, 
such as problems discussed in this paper.

\subsection*{A generic formulation of the optimization problem}
To set the stage, we consider a generic optimization problem $\min_{\sigma}\mathcal{L}(\sigma,\Xi_N)$. Here $\mathcal{L}$ encodes the objective function
(cost) to be minimized. The solutions $\sigma$ lie in some ambient solution space $\Sigma_N$, which is typically very high dimensional, often discrete,
with $N$ encoding its dimension. The space $\Sigma_N$ is equipped with some metric (distance) $\rho_N(\sigma,\tau)$ defined for each pairs of 
solutions $\sigma,\tau\in\Sigma_N$.
For the max-clique problem we can take $\Sigma_N=\{0,1\}^N$. For the Number Partitioning problem we can take
$\Sigma_N=\{-1,1\}^N$. $\Xi_N$ is intended to denote the associated randomness of the problem. So, for example, 
for the max-clique problem, $\Xi_N$ encodes
the random graph $\G(N,1/2)$, and $\sigma\in \{0,1\}^N$ 
encodes a set of nodes constituting a clique, with $\sigma_{i}=1$ if node $i$ is in the clique
and $=0$ otherwise. We denote by $\xi$ an instance generated according to the probability law of $\Xi_N$. In the present context, $\xi$ is any 
graph,  the likelihood of generating of which is $2^{-{N(N-1)\over 2}}$. 
Finally, $-\mathcal{L}(\sigma,\Xi_N)$ is the number of ones in the vector $\sigma$. Now, if $\sigma_{N}$  does not actually encode a clique
(that is for at least one of the edges $(i,j)$ of the graph we have $\sigma_{N,i}=\sigma_{N,j}=1$) we can easily augment the setup by declaring
$\mathcal{L}(\sigma,\Xi_N)=\infty$ for such ''non-clique'' encoding vectors. For the Number Partitioning problem, $\Xi_N$ is just 
the probability distribution associated with an $N$-vector
of independent standard  Gaussians. For each $\sigma\in \Sigma_N=\{-1,1\}^N$ and each instance $\xi$ of $\Xi_N$,
 the value of the associated partition is $\mathcal{L}(\sigma,\xi)=|\langle \sigma,\xi\rangle|$.
Here $\langle x,y\rangle$ denotes an inner product between vectors $x$ and $y$. Thus, for the largest clique problem we have that the random variable
$c^*=-\min_{\sigma\in\Sigma_N}\mathcal{L}(\sigma,\Xi_N)$ is approximately $2\log_2 N$ with high degree of likelihood. For the Number Partitioning
problem we have that $c^*=\min_{\sigma\in\Sigma_N}\mathcal{L}(\sigma,\Xi_N)$ is approximately $\sqrt{N}2^{-N}$ with high degree of likelihood as well.

\subsection*{The OGP and its variants}
\begin{figure}
\begin{center}
\scalebox{.5}{
\includegraphics{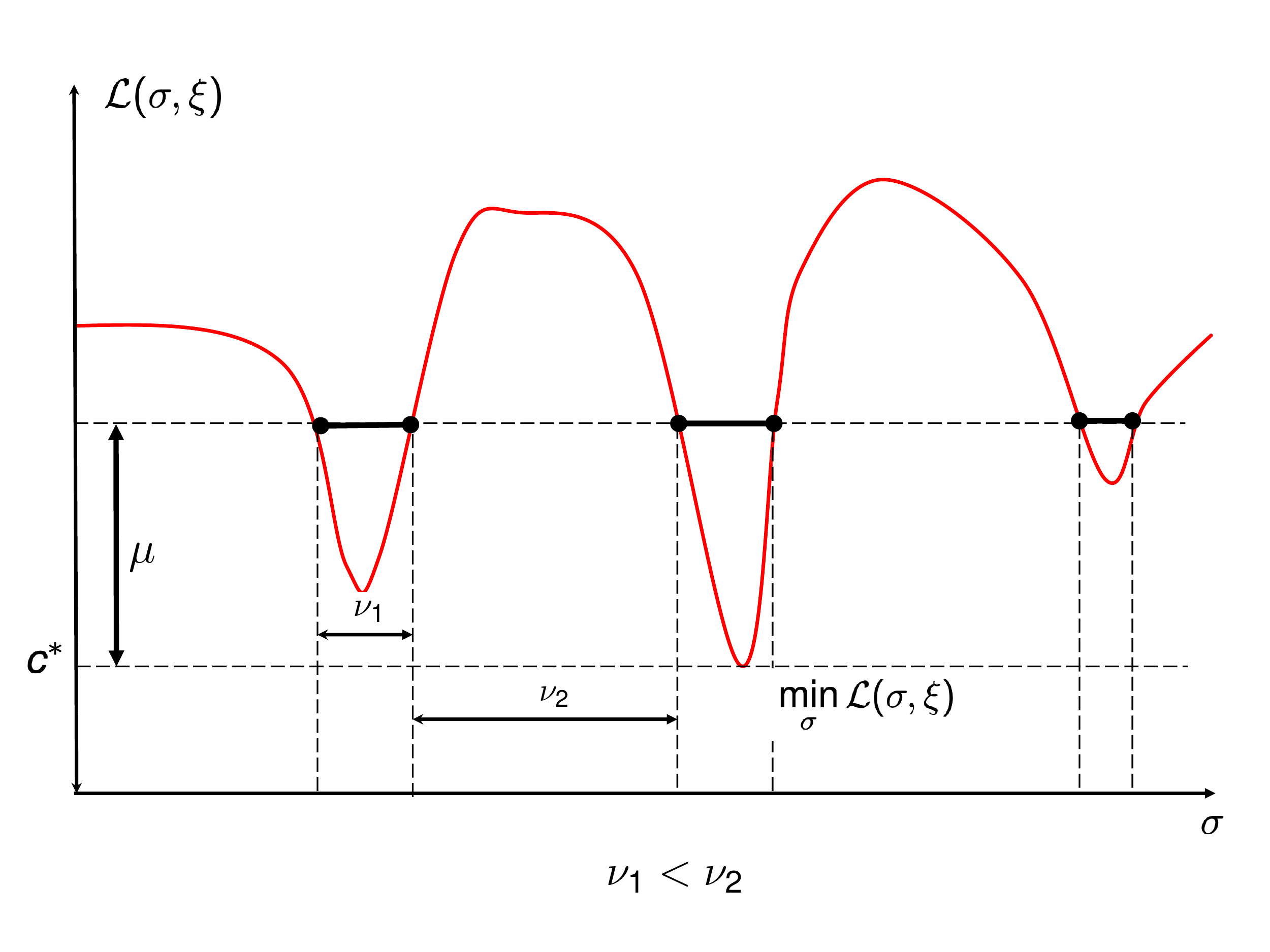}
}
\end{center}
\caption{Landscape exhibiting the OGP: Solutions are split into clusters, with diameter 
of each cluster smaller than the distance between any  pair clusters.}
\label{fig:OGP-2d}
\end{figure}

\begin{figure}
\begin{center}
\scalebox{.5}{
\includegraphics{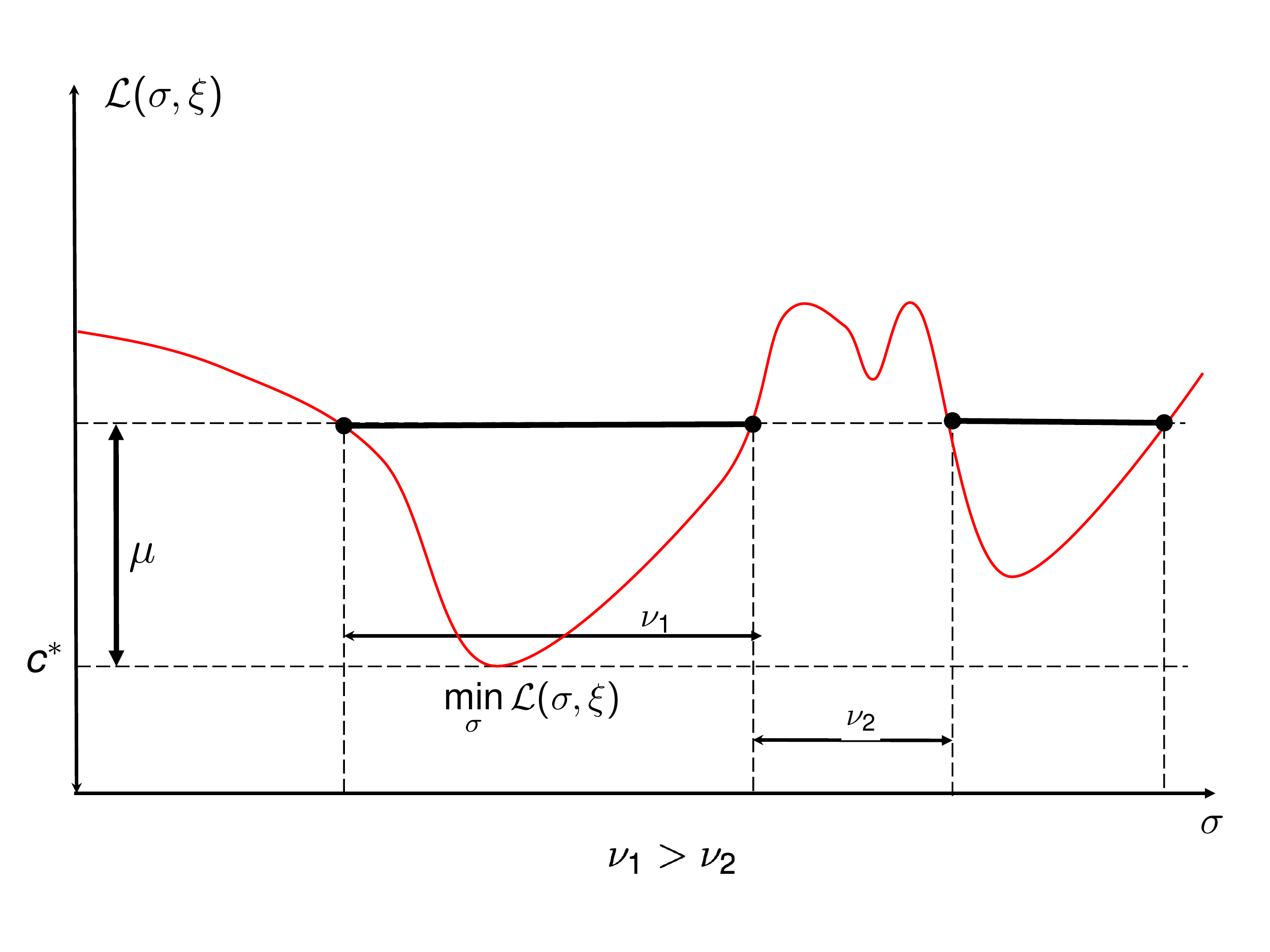}
}
\end{center}
\caption{Landscape not exhibiting the OGP: diameter of  
one cluster is larger than distance between one pair of clusters.}
\label{fig:No-OGP-2d}
\end{figure}

We now introduce the  \emph{Overlap Gap Property }(abbreviated as OGP) and its several variants. 
The term was introduced   in~\cite{gamarnik2018finding}, 
but the property itself was discovered by 
Achlioptas and Ricci-Tersenghi~\cite{achlioptas2006solution}, and 
Mezard, Mora and Zecchina~\cite{mezard2005clustering}.
The definition of the OGP pertains to a particular
instance $\xi$ of the randomness $\Xi_N$. We say that the optimization 
problem $\min_\sigma \mathcal{L}(\sigma,\xi)$ exhibits the OGP with values
$\mu>0, 0\le \nu_1<\nu_2$ if  for every two solutions $\sigma,\tau$ which are $\mu$-optimal in the 
additive sense, namely satisfy 
$\mathcal{L}(\sigma,\xi)\le c^*+\mu,\mathcal{L}(\tau,\xi)\le c^*+\mu$, it is the case that 
either $\rho_N(\sigma,\tau)\le \nu_1$ or 
$\rho_N(\sigma,\tau)\ge \nu_2$. Intuitively, the definition says that every two solutions 
which are ''close''  (within an additive error $\mu$) 
to optimality, are either ''close'' (at most distance $\nu_1$) to each other,  
or ''far'' (at least distance $\nu_2$) from each other, 
thus exhibiting a fundamental topological discontinuity of the set of distances  
of near optimal solutions. In the case
of random instances $\Xi_N$ we say the the problem exhibits the OGP 
if the problem $\min\mathcal{L}(\sigma,\xi)$ exhibits the OGP
with high likelihood when $\xi$ is generated according to the law $\Xi_N$.
An illustration of the OGP is depicted on Figure~\ref{fig:OGP-2d}. 
The notion of the ''overlap'' refers to the fact that distances in 
normed space are directly relatable to inner products, commonly called \emph{overlaps}
in the spin glass theory, via $\|\sigma-\tau\|_2^2=\|\sigma\|_2^2+\|\tau\|_2^2-2\langle \sigma,\tau\rangle$, 
and the fact that solutions $\sigma,\tau$
themselves often have identical or close to identical norms. The OGP is of interest only for 
certain choices of parameters $\mu,\nu_1,\nu_2$ which is 
always problem dependent. Furthermore, all of these parameters along with the optimal
value $c^*$ are usually dependent on the problem size, such as the number of Boolean variables
in the K-SAT model or number of nodes  in a graph. 
In particular, the property  is of interest only if there are 
pairs of $\mu$-optimal solutions $\sigma,\tau$ 
satisfying both $\rho_N(\sigma,\tau)\le \nu_1$ and $\rho_N(\sigma,\tau)\ge \nu_2$ property. 
The first case is trivial as we can take $\tau=\sigma$,
but the existence of pairs with distance at least $\nu_2$ needs to be verified. 
Establishing the OGP rigorously can be either straightforward or technically very involved, 
again depending on the problem. 
It is should not be surprising that the presence of the OGP presents a potential difficulty 
in finding an optimal solution, due to the presence of 
multiple local minima, similarly to the lack of convexity property. 
An important distinction should be noted, however,  between the OGP and the lack of 
convexity property. The function $\mathcal{L}(\sigma,\xi)$
can be  non-convex, but not exhibiting the OGP, as depicted on Figure~\ref{fig:No-OGP-2d}. 
Thus the ''common'' intractability obstruction presented by
non-convexity is not identical with the OGP.  Also, the solution space $\Sigma_N$ is 
often discrete (such as the binary cube) rendering the notions
of convexity non-applicable.

We now extend the OGP definition to the so-called \emph{ensemble} 
Overlap Gap Property, abbreviated as the e-OGP, and  the \emph{multi}-overlap gap 
property, abbreviated as the m-OGP. We say that \emph{a set} of problem 
instances $\Xi$ satisfies the e-OGP with parameters $\mu>0,0\le \nu_1<\nu_2$ if
for every pair of instances $\xi,\psi\in\Xi$, for  every $\mu$-optimal solution $\sigma$ of the $\xi$ instance 
(namely $\mathcal{L}(\sigma,\xi)\le \min_{\sigma'}\mathcal{L}(\sigma',\xi)+\mu$), 
and every $\mu$-optimal solution $\tau$ of the $\psi$ instance
(namely $\mathcal{L}(\tau,\psi)\le \min_{\tau'}\mathcal{L}(\tau',\psi)+\mu$), 
it is the case that either $\rho_N(\sigma,\tau)\le \nu_1$
or $\rho_N(\sigma,\tau)\ge \nu_2$, and in the case when instances $\xi$ and $\psi$ are probabilistically independent,
the former case (i.e. $\rho_N(\sigma,\tau)\le \nu_1$) is not the possible.
 The set $\Xi$  represents a collection of problem instances over which the optimization problem is considered.
For example $\Xi$ might represent a collection of correlated  random graphs $\G(N,1/2)$, or random matrices
or random tensors. Indeed, below we will provide an example of a family
of correlated random graphs $\G(N,1/2)$ as a main 
example of the case when the e-OGP holds. We see that the OGP is the special case of the e-OGP
when $\Xi$ consist of a single instance $\xi$.

Finally, we say that a family of instances $\Xi$ 
satisfies the m-OGP with parameters $\mu,\nu_1<\nu_2$ and $m$, if for every $m$-collection of instances
$\xi_1,\ldots,\xi_m\in\Xi$ and every collection 
of solutions $\sigma_1,\ldots,\sigma_m$ that are  $\mu$-optimal with respect to $\xi_1,\ldots,\xi_m$,
\emph{at least one pair} $\sigma_i,\sigma_j$ 
satisfies $\rho_N(\sigma_i,\sigma_j)\le \nu_1$ or $\rho_N(\sigma_i,\sigma_j)\ge \nu_2$. Informally,
the m-OGP means that one cannot find $m$ 
near optimal solutions for $m$ instances such that all  $m(m-1)/2$ pairs of distances lie in the interval $(\nu_1,\nu_2)$.  
Clearly the case $m=2$ boils down to the e-OGP discussed earlier. 
In many applications  the difference 
between $\nu_1$ and $\nu_2$ is often significantly smaller than the value of $\nu_1$ itself, 
that is $\nu_1\approx \nu_2$. Thus, roughly speaking
the m-OGP means that one cannot find $m$ near optimal solutions so that all pairwise distances are $\approx \nu_1$.
The first usage of OGP as an obstruction to algorithms was adopted by Gamarnik and Sudan 
in~\cite{gamarnik2014limits}. The first application of 
m-OGP as an obstruction to algorithms was adopted by Rahman and Virag in~\cite{rahman2017local}.
While their variant of m-OGP proved useful for some applications
as in Gamarnik and Sudan~\cite{gamarnik2017performance}, 
and Coja-Oghlan, Haqshenas and Hetterich~\cite{coja2017walksat},
since then the definition of m-OGP has been extended
to ''less symmetric'' variants as one by Wein in~\cite{wein2020optimal} and by Bresler and Huang 
in~\cite{bresler2021algorithmic}. We will 
not discuss the nature of these extensions and instead refer to the aforementioned references for further details.
e-OGP was introduced in Chen et al.~\cite{chen2019suboptimality}.

\subsection*{OGP is an obstruction to stability}
Next we discuss how the presence of variants of the  OGP constitutes an obstruction to algorithms. 
The broadest class of algorithms for which such
an implication can be established are   algorithms that are stable with respect to a small perturbation of instances $\xi$. 
An algorithm $\mathcal{A}$ is  viewed here simply as a mapping from instances $\xi$ 
to solutions $\sigma$, which we abbreviate as $\mathcal{A}(\xi)=\sigma$. 
Suppose, we have a parametrized collection of instances $\xi_t$ with discrete parameter $t$ taking 
values in some interval $[0,T]$. 
We say that the algorithm $\mathcal{A}$ is stable or specifically $\kappa$-stable 
with respect to the family $(\xi_t)$ if for every $t$,
$\rho_N(\mathcal{A}(\xi_{t+1}),\mathcal{A}(\xi_{t}))\le \kappa$. Informally, if we think of $\xi_{t+1}$ as an instance  
obtained from $\xi_t$ by a small perturbation,
the output of the algorithm does not change much: the algorithm is not very sensitive to the input. 
Continuous versions of such stability have been considered as well, specifically 
in the context of models with Gaussian distribution. Here $t$ is a continuous parameter,
and the algorithm is stable with sensitivity value $\delta$ if for every $t$
$\rho_N(\mathcal{A}(\xi_{t+\delta}),\mathcal{A}(\xi_{t}))\le \kappa$. Often these bounds
are established only with probabilistic guarantees, both in the case of discrete and continuously valued $t$.

Now assume that the e-OGP holds for a family of instances $\xi_t, t\in [0,T]$.  
Assume two additional conditions: (a)  the regime
$\rho_N(\sigma,\tau)\le \nu_1$ for $\mu$-optimal solutions $\sigma$ of $\xi_0$, and $\mu$-optimal solutions 
$\tau$ of $\xi_T$ is non-existent (namely every $\mu$-optimal solution of $\xi_0$ 
is ''far'' from every $\mu$-optimal solution of $\xi_T)$; and 
(b) $\kappa<\nu_2-\nu_1$. The property (a) is typically verified since
the end points $\xi_0$ and $\xi_T$ of the interpolated sequence are often independent,
and (a) can be checked by straightforward moment arguments. The verification of condition (b)
typically involves technical and problem dependent arguments. 
The conditions (a) and (b) above plus the OGP allow us to conclude that 
any $\kappa$-stable algorithm  fails to produce a $\mu$-optimal solution. This is seen by
a simple contradiction argument based on continuity:  
consider the sequence of solutions $\sigma_t=\mathcal{A}(\xi_t), t=0,\ldots,T$, produced
by the algorithm $\mathcal{A}$. We have
$\rho_N(\sigma_t,\sigma_{t+1})\le \kappa$ for every $t$ 
and $\rho_N(\sigma_0,\sigma_T)\ge \nu_2$ by the OGP and assumption (a) above. Suppose, for the purposes of contradiction,  
that every solution $\sigma_t$
produced by the algorithm is $\mu$-optimal  for 
every instance $\xi_t$. Then by the OGP it is the case that, in particular,  
$\rho_N(\sigma_0,\sigma_t)$ is either at most $\nu_1$ or at least $\nu_2$ for every $t$. Since 
$\rho_N(\sigma_0,\sigma_T)\ge \nu_2$, there exists $t$, possibly $t=0$, such that
$\rho_N(\sigma_0,\sigma_t)\le \nu_1$ and $\rho_N(\sigma_0,\sigma_{t+1})\ge \nu_2$.
Then 
$\kappa\ge \rho_N(\sigma_t,\sigma_{t+1})\ge |\rho_N(\sigma_0,\sigma_t)-\rho_N(\sigma_0,\sigma_{t+1})|\ge \nu_2-\nu_1$,
which is a contradiction with the assumption (b).

\begin{figure}
\begin{center}
\scalebox{.35}{\includegraphics{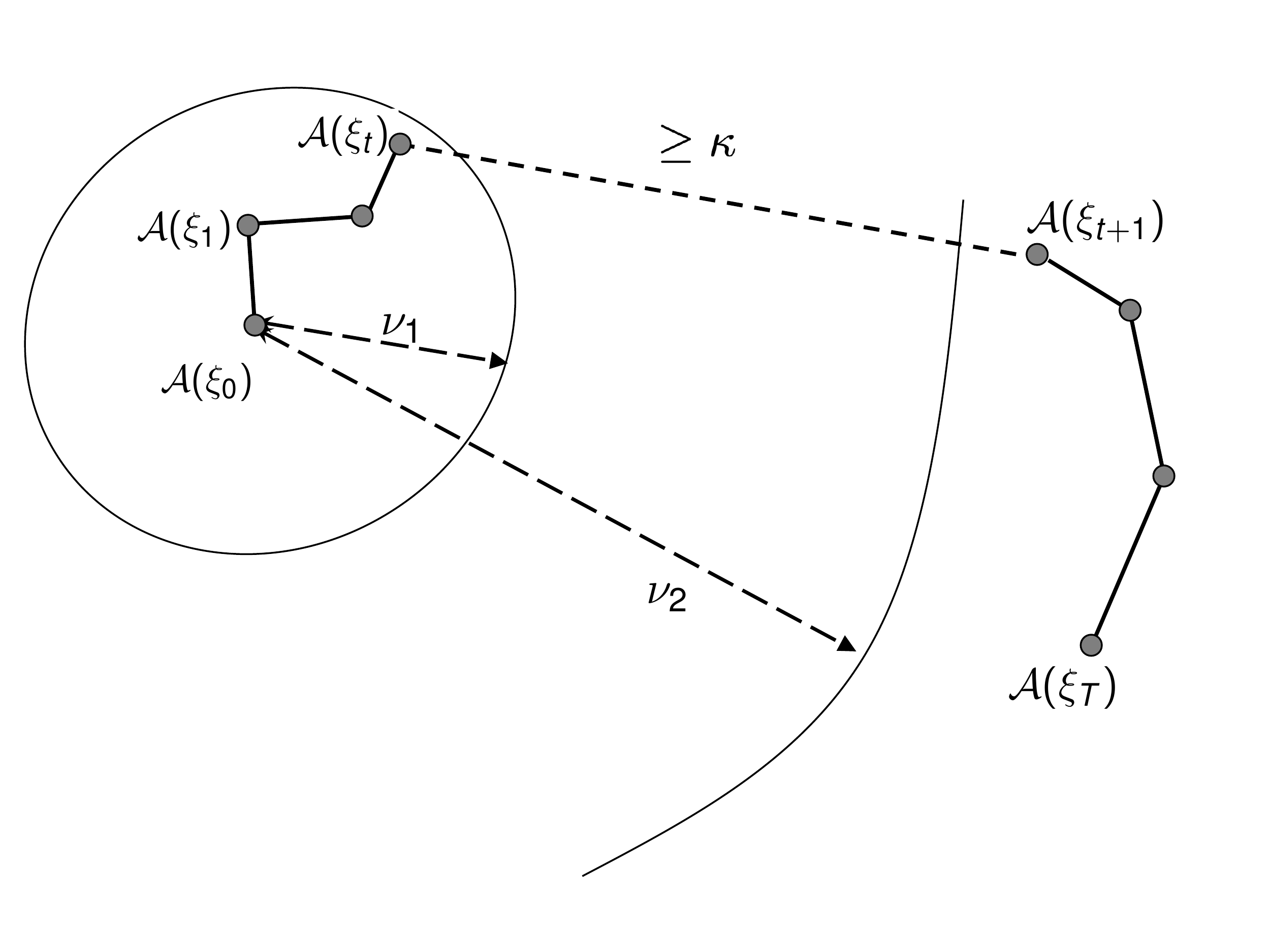}}
\end{center}
\caption{The smaller circle represents $\mu$-optimal solutions at distance $\le \nu_1$
from $\mathcal{A}(\xi_0)$. The complement to the larger circle 
represents $\mu$-optimal solutions at distance $\ge \nu_2$
from $\mathcal{A}(\xi_0)$. As distance between the circle boundaries is $\nu_2-\nu_1>\kappa$,
at some instance $t$ the distance between successive solutions 
$\mathcal{A}(\xi_t)$ and $\mathcal{A}(\xi_{t+1})$ has to exceed $\kappa$.
}
\label{fig:Stable-algorithms}
\end{figure}

Simply put, the crux of the argument is that the algorithm cannot ''jump'' over the gap $\nu_2-\nu_1$, 
since the incremental distances, bounded by $\kappa$
are two small to allow for it.
This is the main method by which stable algorithms are ruled out in presence of the OGP and  
is illustrated on 
Figure~\ref{fig:Stable-algorithms}. On this figure the smaller circle  
represent all of the $\mu$-optimal solutions which are at most distance $\nu_1$ from $\sigma_0$,
across all instances $t\in [0,T]$. The complement to the larger circle  
represents all of the $\mu$-optimal solutions which are  \emph{at least} distance $\nu_2$ from $\sigma_0$,
again across all instances $t\in [0,T]$. All $\mu$-optimal solutions across all instances should fall into one of 
these two regions, according to the e-OGP. 
As the sequence
of solutions $\mathcal{A}(\xi_t), t\in [0,T]$ has to cross between these two  regions, at some point $t$ the 
distance between $\mathcal{A}(\xi_t)$ and $\mathcal{A}(\xi_{t+1})$ will
have to exceed $\kappa$, contradicting the stability property.

\begin{figure}
\begin{center}
\scalebox{.5}{
\includegraphics{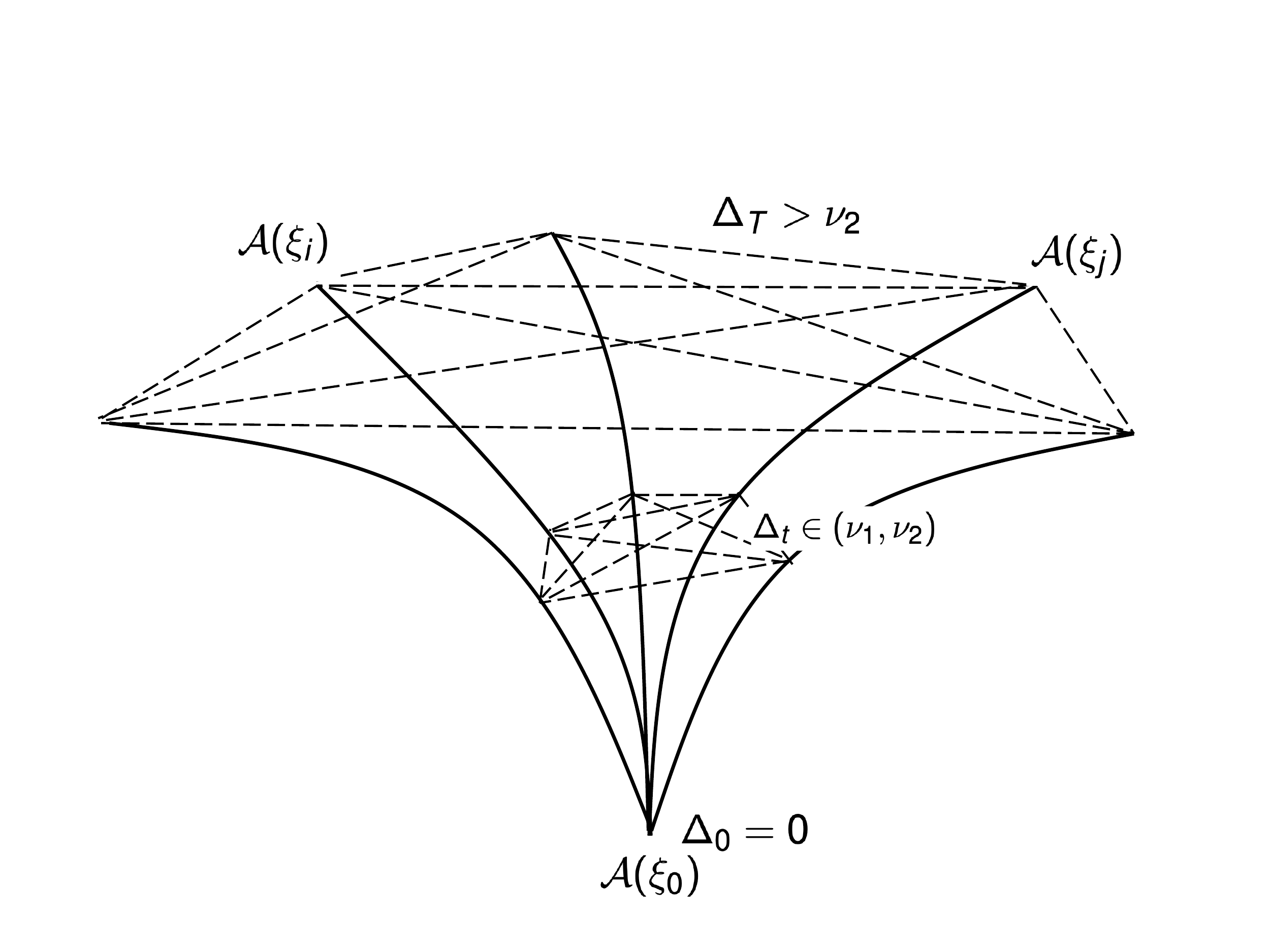}
}
\end{center}
\caption{Pairwise distances $\Delta_t$ are nearly 
identical for all pairs and continuously evolve from $0$ as $t$ increases.
At some time instance $t$, the value of $\Delta_t$ hits the gap region $(\nu_1,\nu_2)$,
implying the same for all ${m\choose 2}$ pairwise distances.}
\label{fig:m-OGP}
\end{figure}

Suppose the model does not exhibit the OGP, but does exhibit the m-OGP (ensemble version). We now describe
how this can also be used as an obstruction to stable algorithms. As above, the argument is based on first
verifying that for two independent instance $\xi$ and $\tilde\xi$ the regime $\rho_N(\sigma,\tau)\le \nu_1$
is not possible with overwhelming probability. Consider $m+1$ independent instances $\xi_0,\xi_1,\ldots,\xi_m$
and consider an interpolated sequence $\xi_{1,t},\ldots,\xi_{m,t}, t\in [0,T]$ with the property
that $\xi_{i,T}=\xi_i$ and $\xi_{i,0}=\xi_0$ for all $i=1,2,\ldots,m$. In other words the sequence starts with $m$
identical copies of instance $\xi_0$ and slowly interpolates towards $m$ instance $\xi_1,\ldots,\xi_m$.
The details of such a construction are usually guided by  concrete problems. Typically, such constructions
are also symmetric in distribution, so that, in particular, all pairwise expected distances between the algorithmic
outputs $\E{\rho_{N}\left(\mathcal{A}(\xi_{i,t}),\mathcal{A}(\xi_{j,t})\right)}$ are the same, say denoted
by $\Delta_t, t\in [0,T]$.
Furthermore, in some cases a concentration around the expectation can be also established. As an implication
this set of identical pairwise distances $\Delta_t$ spans values from $0$ to value larger than $\nu_1$ 
by the assumption (a).
But the stability of the algorithm $\mathcal{A}$ also implies that at some point $t$  it will be the case
that $\Delta_t\in (\nu_1,\nu_2)$, contradicting the m-OGP. This is illustrated on Figure~\ref{fig:m-OGP}.
The high concentration property discussed above was established using standard methods 
in~\cite{gamarnik2017performance} and~\cite{wein2020optimal}, but the approach in~\cite{GamarnikKizildagNPP}
was based on a different  arguments employing methods from Ramsey extremal combinatorics field. Here
the idea is to generate many more $M\gg m$ independent instances than the value of $m$ arising in m-OGP
and showing using Ramsey theory results that there should exist a subset of $m$-instances such that
all pairwise distances fall into interval $(\nu_1,\nu_2)$.

\subsection*{OGP for concrete models}
We now illustrate the OGP and its variants for the Number Partitioning problem and the maximum clique problem as  examples. 
The discussion will be informal and references containing detailed derivations will be provided.
The instance
$\xi$ of the Number Partitioning problem is a sequence $X_1,\ldots,X_N$ of independent standard normal random variables. 
For every binary vector $\sigma\in \{-1,1\}^N$, the inner product $Z(\sigma)=\langle \xi, \sigma\rangle$ 
is a mean zero normal random variable 
with variance $N$. The likelihood that $Z(\sigma)$ takes value in an interval around zero with length 
$\sqrt{N}2^{-N}$ is roughly $(2\pi)^{-{1\over 2}}$ times the length of the interval, namely times $2^{-N}$. Thus the expected
number of such vectors $\sigma$ is of constant  order $(2\pi)^{-{1\over 2}}$ 
since there are $2^N$ choices for $\sigma$. This gives
an intuition for why there do exist vectors achieving value $\sqrt{N}2^{-N}$. Details can be found  
in \cite{karmarkar1982differencing}. Any choice of order magnitude smaller
length interval leads to expectation converging to zero and thus 
non-achievable with overwhelming probability. Now, to discuss the OGP,
fix constants $\alpha,\rho\in (0,1)$ and consider pairs of 
vectors $\sigma,\tau$ which achieve objective value $\sqrt{N}2^{-\alpha N}$
and have the scaled inner product $N^{-1}\langle \sigma,\tau\rangle =\rho$. The expected number of such pairs is roughly 
${N\choose {1-\rho \over 2} N}$ times the area of the square with side length approximately $2^{-\alpha N}$, namely
$2^{-2\alpha N}$. This is because
the likelihood that a pair of correlated gaussians $Z(\sigma),Z(\tau)$ falls into a very small square around zero is dominated
by the area of this square up to a constant. 
Approximating  ${N\choose {1-\rho \over 2} N}$ as $\exp\left(N H({1-\rho\over 2})\right)$, where
$H(x)=-x\log x-(1-x)\log (1-x)$ is the standard binary entropy, the expectation of the number of such pairs evaluates
to  roughly 
$\exp\left(N\left(H({1-\rho\over 2})-2\alpha\log 2 \right)\right)$. As the largest value of $H(\cdot)$ is $\log 2$, we obtain
an important discovery: for every $\alpha\in (1/2,1)$ 
there is a region of $\rho$ of the form $(\rho_0,1)$ for which the expectation
is exponentially small in $N$, and thus the value $\sqrt{N}2^{-\alpha N}$ is not achievable 
by pairs of  partitions $\sigma$ and $\tau$ with scaled inner products in $(\rho_0,1)$,  
with 
overwhelming probability. Specifically, for every pair of $\mu$-optimal 
solutions $\sigma,\tau$ with $\mu=(2^{-\alpha N}-2^{-N})$, 
it is the case that $N^{-1}\langle \sigma,\tau\rangle$ is either $1$
or at most $\rho_0$. With this value of $\mu$ and $\nu_1=0,\nu_2=N(1-\rho_0)/2$, we conclude that the model exhibits the OGP. 

Showing the  ensemble version of OGP (namely e-OGP) can be done using very 
similar calculations, but what about  values of $\alpha$ smaller than $1/2$?
After all, the state of the art algorithms achieve only order $2^{-O(\log^2 N)}$ which corresponds to $\alpha=0$ at scale.
It turns out \cite{GamarnikKizildagNPP} that in fact the m-OGP holds for this 
model for all strictly positive $\alpha$ with a judicious
choice of constant value $m$. Furthermore, the m-OGP extends all the 
way to $2^{-O(\sqrt{N \log N})}$ by taking $m$ growing with $N$, but not
beyond that value, at least using the available proof techniques. 
Thus, at this stage we may conjecture that the problem is algorithmically hard for objective values smaller than order
$2^{-O(\sqrt{N\log N})}$, but beyond this value we don't have a plausible predictions either for hardness or tractability.
In conclusion, 
at least at the exponential scale $\exp(-\Theta(N))$ 
we see that the OGP gives a theory of algorithmic hardness consistent
with the state of the art algorithms.

The largest clique problem is another example where the OGP based predictions 
are consistent with the state of the art algorithms. 
Recall from the earlier section,  that while the largest clique in $\G(N,1/2)$ 
is roughly $2\log_2 N$, the largest size clique achievable by known algorithm
is roughly half-optimal with size approximately $\log_2 N$. Fixing $\alpha\in (1,2)$ 
consider pairs of cliques $\sigma,\tau$ with sizes
roughly $\alpha\log_2 N$. This range of $\alpha$ corresponds to clique sizes known 
to be present in the graph existentially with overwhelming
probability, but not achievable algorithmically. As it turns out, 
see \cite{gamarnik2014limits},\cite{gamarnik2020lowFOCS}, the model
exhibits the OGP for $\alpha\in (1+1/\sqrt{2},2)$ and overlap 
parameters $\nu_1(\alpha)<\nu_2(\alpha)$, both of which are at the scale $\log_2 N$.
Specifically, every two cliques of size $\alpha\log_2 N$ have intersection size either at 
least $\tilde\nu_1(\alpha)$ or at most  $\tilde\nu_2(\alpha)$, for some values $\tilde\nu_1(\alpha)>\tilde\nu_2(\alpha)$,
both at scale $\log_2 n$. This easily implies OGP with $\nu_j(\alpha)=2\alpha-\tilde\nu_j(\alpha), j=1,2$.
The calculations in~\cite{gamarnik2014limits} and~\cite{gamarnik2020lowFOCS} were carried out 
for a similar problem of finding a largest independent set (a set with no internal edges) in sparse
random graphs $G(N,p_N)$ with $p_N$ of the order $O(1/N)$, but the calculation details for the dense graph $\G(N,1/2)$
are almost identical. 
The end result
is that $\tilde\nu_j(\alpha)=x_j(\alpha)\log_2 N$, where $x_1(\alpha)$ and $x_2(\alpha)$ 
are the two roots of the quadratic equation $x^2/2-x+2\alpha-\alpha^2=0$, namely $1\pm\sqrt{1-2(2\alpha-\alpha^2)}$,
and roots exist if and only if $\alpha>1+1/\sqrt{2}$. For recent algorithmic results for the related 
problem on random regular
graphs with small degree see~\cite{marino2020large}.

\begin{figure}
\begin{center}
\scalebox{.5}{\includegraphics{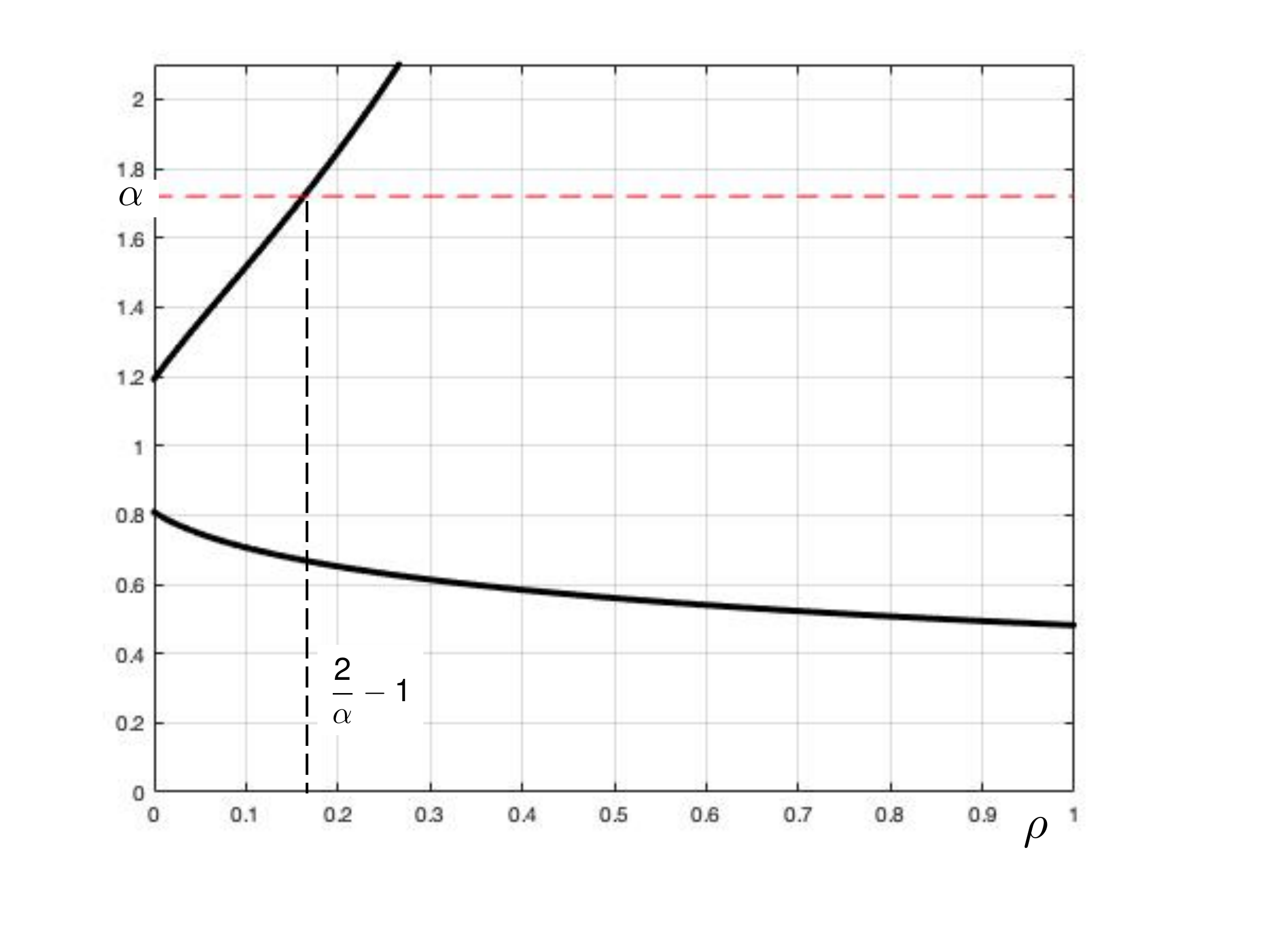}}
\end{center}
\caption{Overlap gap as a function of correlation $\rho$. Realizable overlaps are above the top curve and below the bottom curve only. 
To the right of $\rho^*=2/\alpha-1=2/1.72-1\approx 0.163$ only values below the bottom curve correspond to the realizable overlaps.
}
\label{fig:OGP-Cliques}
\end{figure}

To illustrate the ensemble version of the OGP,  consider an ensemble $G_1,\ldots,G_{N\choose 2}$ 
of random $\G(N,1/2)$ graphs, 
constructed as follows. Generate $G_0$ according to the probability law of $\G(N,1/2)$. 
Introduce a random uniform order $\pi$ of ${N\choose 2}$ pairs 
of nodes $1,\ldots,N$ and generate graph $G_{t}$ from $G_{t-1}$ by resampling
the pair $(i_{t},j_{t})$ -- the $t$-th pair in the order $\pi$, 
and leaving other edges of $G_{t-1}$ intact. This way each graph $G_t$ individually
has the law of $\G(N,1/2)$, but neighboring graphs $G_{t-1}$ 
and $G_t$ differ by at most one edge. The beginning and the end graphs $G_0$ and $G_{N\choose 2}$
are clearly independent. Now fix $\alpha\in (1+1/\sqrt{2},2)$ and $x\in (0,\alpha)$
A simple calculation shows that the expected number of $\alpha\log_2 N$ size cliques $\sigma$ in $G_{t_1}$ and
same size cliques $\tau$ in $G_{t_2}$, such that the size of the intersection of $\sigma$ and $\tau$ is $x\log_2 N$ is 
approximately given by $\exp\left((1+o(1))\log_2^2 N\left((1-\rho) x^2/2-x+2\alpha-\alpha^2\right)\right)$, 
where $\rho=(t_2-t_1)/{N\choose 2}$
is the (rescaled) number of edges of $G_{t_2}$ which were resampled from $G_{t_1}$. 
Easy calculation shows that the function $(1-\rho) x^2/2-x+2\alpha-\alpha^2$
has two roots $x_1>x_2$ in the interval $(0,\alpha)$ 
given by $(1-\rho)^{-1}\left(1\pm\sqrt{1-2(1-\rho)(2\alpha-\alpha^2)}\right)$,
when 
the larger root is smaller than $\alpha$, namely when $\rho<{2\over \alpha}-1$.
On the other hand, when $\rho>{2\over \alpha}-1$, including
the extreme case $\rho=1$ corresponding to $t_1=0, t_2={N\choose 2}$, there is only one root in $(0,\alpha)$. 
The model thus not only does exhibit the e-OGP  for the ensemble $G_t, 0\le t\le {N\choose 2}$. Additionally
the value
$\rho^*={2\over \alpha}-1$ represents a new type of phase transition, not known earlier to the best of the author's knowledge. 
When $\rho<\rho^*$ 
every two  cliques with size $\alpha\log_2 N$ 
have intersection  size either at least  $x_1\log_2 N$ or at most 
$x_2\log_2 N$. On the other hand, when $\rho>\rho^*$, the second
regime disappears, and any two
such cliques have intersection  size only at most  $x_2\log_2 N$. 
The situation is illustrated on the Figure~\ref{fig:OGP-Cliques}
for the special case $\alpha = 1.72$ and $\rho^*=2/\alpha-1=0.163..~$. 
This phase transition is related to the so-called chaos property of spin glasses which we discuss
below, but has qualitatively different behavior. In particular, the analogous parameter $\rho^*$
in the context of spin glass models has value $\rho^*=0$. More on this below.

The value $\alpha=1+1/\sqrt{2}$ is not tight with respect to the 
apparent algorithmic hardness which occurs at $\alpha=1$ as discussed earlier.
This is where the multi-OGP comes to rescue. A variant of an (assymetric) m-OGP has 
been established recently in~\cite{wein2020optimal}, building
on an earlier symmetric version of m-OGP, discovered in~\cite{rahman2017local}, 
ruling out the class of stable algorithms as potential contenders for solving
the Independent Set problem above $\alpha=1$. More specifically, in both works
(for the case of local algorithms in~\cite{rahman2017local} and for the case of 
low-degree polynomials based algorithms in~\cite{wein2020optimal}) it is shown 
that for every fixed $\epsilon>0$ there exists a constant $m=m(\epsilon)$ such that the 
model exhibits m-OGP with this value of $m$ for independent sets corresponding to $\alpha=1+\epsilon$.
While, it has not be formally verified, it appears based on the first moment argument 
that value $m$ growing with $\epsilon$ is needed, 
in the sense that \emph{no fixed} value of $m$ achieves the OGP all the way down to $\alpha=1$, 
but constant values of $m$ which are $\epsilon$-dependent do.
It is interesting to contrast this with the Number Partitioning problem where it was crucial
to take $m$ growing with $N$ in order the bridge the gap between the algorithmic and existential 
thresholds.

\subsection*{Stability of low-degree polynomials}
How does one establish the link of the form \emph{OGP implies Class $\mathcal{C}$ of algorithms fails}, and what classes $\mathcal{C}$ of algorithms can
be proven to fail in presence of the OGP? As discussed earlier, the failure is established by showing \emph{stability} (insensitivity)
of the algorithms in class \emph{C} to the changes of the input. This stability forces the algorithm to produce pairs of outputs with
overlap falling into a region ''forbidden'' by the OGP. The largest class of algorithms to the day which is shown to be stable 
is the class of algorithms informally dubbed as ''low-degree polynomials''. Roughly speaking this is a class of algorithms where the solution
$\sigma\in\Sigma_N$ is obtained by constructing an $N$-sequence of multi-variate polynomials $p_j(\xi), j=1,\ldots,N$ with variables
evaluated at entries of the instance $\xi\in\Xi_N$. The largest degree $d$ of the polynomial is supposed to be low, though depending on the
problem it can take even significantly high value. Algorithms based on low-degree polynomials gained a lot of prominence recently
due to connection with the so-called Sum-of-Squares method and represent arguably the strongest known class of polynomial time algorithms
\cite{hopkins2017efficient},\cite{hopkins2017power},\cite{hopkins2018statistical}. 
What can be said about this method in the context of the problems exhibiting the OGP? 
It was shown in~\cite{gamarnik2020lowFOCS},\cite{wein2020optimal} and~\cite{bresler2021algorithmic} 
that degree-$d$ polynomial based algorithms
are stable for the problems of finding a ground state of a p-spin model,  
finding a large independent set of a sparse \ER graph $\G(N,c/N)$, and for the problem of finding
a satisfying assignment of a random K-SAT problem.
This is true 
even when $d$ is as large as $O(N/\log N)$. Thus algorithms based on polynomials even with degree $O(N/\log N)$ fail to find near optimum
solutions for these problems due to the presence of the OGP. Many  algorithms can be modeled as special cases
of low degree polynomials, and therefore ruled out by OGP, 
including the so-called Approximate Message Passing algorithms~\cite{gamarnik2019overlapAoP},\cite{gamarnik2020lowFOCS},
local algorithms considered in~\cite{gamarnik2014limits},\cite{rahman2017local} and even quantum versions of local algorithms known
as Quantum Approximate Optimization Algorithm (QAOA)~\cite{farhi2020quantumRandom}.

\subsection*{OGP and the problem of finding ground states of p-spin models}
In addition to the maximum clique problem, there is another large class of optimization 
problem for which the OGP approach provides a tight
classification of solvable vs not yet solvable in polynomial time problems. 
It is the problem of finding ground states of  p-spin models -- the
very class of optimization problems which led to studying the overlap distribution 
and overlap gap type properties to begin with. The problem is described
as follows: suppose one is given a $p$-order tensor $J$ of side length $N$. 
Namely, $J$ is  described as a collection of real values $J_{i_1,\ldots,i_p}$
where $i_1,\ldots,i_p$ range between $1$ and $N$. Given the solution space $\Sigma_N$, 
the goal is finding a solution $\sigma\in\Sigma_N$
(called ground state) which
minimizes the inner product 
$\langle J,\sigma\rangle =\sum_{i_1,\ldots,i_p}J_{i_1,\ldots,i_p}\sigma_{i_1}\cdots \sigma_{i_p}$.
 The randomness
in the model is derived from the randomness of the entries of $J$. A common assumption, 
capturing most of the structural and algorithmic complexity 
of the model, is that the entries $J_{i_1,\ldots,i_p}$ are independent standard normal entries.
Two important
special cases of the solution space $\Sigma_N$ include the case when $\Sigma_N$ 
is a binary cube $\{-1,1\}^N$, and when it is a sphere of a fixed radius (say $1$) in the 
$N$-dimensional real space, the latter case being referred to as \emph{spherical} spin glass model. 
The special case $p=2$ and $\Sigma_N=\{-1,1\}^N$
is the celebrated Sherrington-Kirkpatrick model \cite{sherrington1975solvable} -- the centerpiece of the spin glass theory.
The problem of computing the value of this
optimization problem above was at the core of the spin glass theory in the past four decades, 
which led to the replica symmetry breaking technique
first introduced in the physics literature by Parisi~\cite{parisi1980sequence}, 
and then followed by a series of mathematically rigorous 
developments, including those by Talagrad~\cite{TalagrandBook} and Panchenko~\cite{panchenko2013sherrington}.

The algorithmic progress on this problem, however, is quite recent. 
While the power of the Monte Carlo based simulation results have been tested numerous
times, the first rigorous progress has appeared only in the past few years. 
In the important development by Subag~\cite{subag2018following}
and Montanari~\cite{montanari2019optimization}  a polynomial time algorithms were constructed for finding
near ground states in these models. Remarkably, this was done precisely 
for regimes where the model \emph{does not} exhibit the OGP as per 
a detailed analysis of the  so-called Parisi measure of the associated Gibbs distribution. A follow-up
development in~\cite{alaoui2020optimization} extended this to models which do exhibit OGP, 
but in this case near ground states are not reached. Instead,
one constructs the best solution within the part of the solution space \emph{not exhibiting} the OGP. 
To put it differently, the algorithms
produce a $\mu$-optimal solution for the smallest value of $\mu$ for which the OGP does not hold. 
Conversely, as shown in a series of recent 
developments~\cite{gamarnik2019overlapAoP},\cite{gamarnik2020low},\cite{sellke2020approximate}, 
the optimization problem problem associated
with finding near ground states cannot be solved within some classes of  algorithms, the largest of which is the class of
algorithms based on low-degree polynomials. The failure of this class of algorithms 
was established exactly along the lines described above:
verifying stability of the  algorithms of interest, and verifying the presence of OGP 
which presents an obstruction to stability.
Interestingly, a class of algorithms, which is not described by low-degree polynomials but which 
is still ruled out by the OGP is the Langevin dynamics 
at a linear in $N$ scale, as shown in~\cite{gamarnik2020low}. The Langevin dynamics is continuous 
version of the Markov Chain Monte Carlo type algorithm. Its stability stems
from the Gronwald's type bound stating that two time dependent Langevin 
trajectories $\sigma(\xi,s),\sigma(\tilde\xi,s), s\in\R_+$
associated with the Langevin dynamics run on two instances $\xi$ and $\tilde\xi$, 
diverge at the rate order $\|\xi-\tilde\xi\|\exp(\kappa s)$ for 
some constant $\kappa>0$. Thus when $s$ is of order $N$ and $\xi$ and $\tilde\xi$ are at distance which
is exponentially small in $N$, one can control this divergence. One then uses e-OGP for an interpolated
sequence $\xi_t, t\in [0,T]$ with discrete values of $t$ with $\exp(-\Theta(N))$ increments to argue stability.
The approach  fails at time scales larger than $N$, since the required concentration bounds underlying
OGP hold only modulo exponentially in $N$ small probability, and as a result the union bound over more than
$\exp(\Theta(N))$ terms is no longer effective. It is a very plausible conjecture, though, that the Langevin
dynamics indeed still fails unless it is ran for exponentially long $s=\exp(\Theta(N))$ time. We leave 
this conjecture as  an interesting open problem.

The analogue of the phase transition $\rho^*={2\over \alpha}-1$ discussed in the context 
of the largest clique problem takes place in the p-spin model as well
in the more extreme form called the \emph{chaos} property: any non-trivial perturbation 
of the instance $J$ whereby any fixed   positive fraction of entries
of $J$ are flipped, results in the instance $\tilde J$ for which every near ground state is nearly 
orthogonal to every near ground state of $J$. To put
it differently $\rho^*$ is effectively zero for this model. This has been established in a series of 
works~\cite{chatterjee2009disorder},\cite{chen2018energy},\cite{eldan2020simple}. 
Notably, the chaos property
itself is not a barrier to polynomial time algorithms since the Sherrington-Kirkpatrick model 
is known to exhibit the chaos property, yet, 
it is amenable to poly-time algorithms as in~\cite{subag2018following} and~\cite{montanari2019optimization}.
The chaos property however is used as a tool to establish the ensemble version of the OGP (e-OGP) as
was done in~\cite{chen2019suboptimality} and~\cite{gamarnik2019overlapAoP}.


\subsection*{OGP, the clustering property, and the curious case of the perceptron model} 
We now return to the subject of weak and strong clustering property, and discuss it now from the perspective
of the OGP and the implied algorithmic hardness. The relationship is interesting,  non-trivial
and perhaps best exemplified by yet another model which exhibits statistical to computation gap: the binary
perceptron model. The input to the model is $M\times N$ matrix of i.i.d. standard normal entries 
$X_{ij}, 1\le i\le M, 1\le j\le N$. A fixed parameter $\kappa>0$ is fixed. The goal is to find 
a binary vector $\sigma\in \{\pm 1\}^N$ such that for each $i, \sum_{1\le j\le N} X_{ij}\sigma_j\in [-\kappa,\kappa]$. 
In vector  notation we need to find $\sigma$ such that $\|X\sigma\|_{\infty}\le \kappa$, where 
$\|x\|_\infty=\max_j |x_j|$ for any vector $x\in\R^N$. This is also known as binary \emph{symmetric}
perceptron model, to contrast it with assymetric model, where the requirement is that $X\sigma\ge \kappa$,
for some fixed parameter $\kappa$ (positive or negative), with inequalities
interpreted coordinate-wise. In the assymetric case the a typical choice is $\kappa=0$. Replica symmetry
breaking based heurstic methods developed in this context by Krauth and Mezard~\cite{krauth1989storage} 
predict a sharp threshold at $\alpha_{\rm SAT}=0.83$ for the case $\kappa=0$, so that when $\alpha<\alpha_{\rm SAT}$
and $M\le \alpha N$ such a vector $\sigma$ exists, and when $\alpha>\alpha_{\rm SAT}$ 
such a vector $\sigma$ does not exist, both with high probability as $N\to\infty$. This was 
confirmed rigorously as an upper bound in~\cite{ding2019capacity}. For the symmetric
case a similar sharp threshold $\alpha_{\rm SAT}(\kappa)$ 
is known rigorously as a $\kappa$ dependent constant $\alpha(\kappa)$,
as established in~\cite{perkins2021frozen} and~\cite{abbe2021proof}.

The known 
algorithmic results however are significantly weaker and scarce. The algorithm by Kim and Rouche~\cite{kim1998covering}
finds $\sigma$ for the assymetric $\kappa=0$ case when $\alpha<0.005$. While the algorithm has not been
adopted to the symmetric case, it is quite likely that a version of it should work here as well for some
positive sufficiently small $\kappa$-dependent constant $\alpha$. Heuristically, the message passing
algorithm was found to be effective at small positive densities, as reported in~\cite{braunstein2006learning}. 
Curiously, though, it is known that the symmetric model exhibits the weak clustering property
\emph{at all} positive densities $\alpha>0$, as was rigorously verified recently in~\cite{perkins2021frozen}
and~\cite{abbe2021proof}. Furthermore, quite remarkably, each cluster consist of singletons! Namely
the cardinality of each cluster is one, akin to ''needles in the haystack'' metaphor. 
A similar picture was established in the context of the 
Number Partitioning problem~\cite{GamarnikKizildagNPP}. This interesting entropic (due to the subextensive
cardinality of clusters) phenomena is fairly novel and its algorithmic implications have been
also discussed in~\cite{bellitti2021entropic} in the context of another constraint satisfaction 
problem -- the random K-XOR model. 
Thus at this stage, assuming the extendability of the Kim-Roche algorithm
and/or validity of the message passing algorithm, we have an example of a model exhibiting weak clustering
property, but   amenable to polynomial time algorithms. This stands in  
in contrast to the random K-SAT model and in fact 
all other models known to the author. As an explanation of this phenomena, the aforementioned 
references  point out that
to the  ''non-uniformity'' in algorithmic choices of the solutions allowing it somehow to bypass
the overwhelming clustering picture exhibited by the weak clustering property.

What about the strong clustering property and the OGP? First let us elucidate the relationship between the two
properties. The pair-wise ($m=2$) OGP which holds for a single instance clearly implies the strong clustering
property: if every pair of $\mu$-optimal solutions is at distance at most $\nu_1$ or at least $\nu_2>\nu_1$,
and the second case is non-vacuous, clearly the set of solutions is strongly clustered. It furthermore,
indicates that the diameter of each cluster is strictly smaller (at scale) than the distances between the clusters,
see Figures~\ref{fig:OGP-2d},\ref{fig:No-OGP-2d}.
In fact this is precisely how the strong clustering property was discovered to begin 
with in~\cite{achlioptas2006solution} and~\cite{mezard2005clustering}. However, 
the converse does not need to be the case,
as one could not rule out an example of a strong clustering property where the cluster
diameters are larger
than the distances between them, and, as a result, 
the overlaps of pairs of solutions span a continuous intervals. At this
stage we are not aware of any such examples.

Back to the symmetric perceptron model, it is known (rigorously, based on the moment computation), that
it does exhibit the OGP, at a value $\alpha$ strictly smaller than the critical 
threshold $\alpha_{\rm SAT}(\kappa)$~\cite{baldassi2020clustering}. As a result
the strong clustering property holds as well above this $\alpha$. Naturally, we conjecture that the problem
of finding a solution $\sigma$ is hard in this regime, and it is very likely that classes of algorithms,
such as low-degree polynomial based algorithm and their implications can be ruled out by the OGP method
along the lines discussed above.

Finally, as the strong clustering property appears to be the most closely related to the OGP, it raises a question
as to whether it can be used as a ''signature'' of hardness in lieu of OGP and/or whether it can 
be used as a method to rule out classes of algorithm, similarly to the OGP based method.
There are two challenges associated with this proposal. First we are not aware of any effective method
of establishing the strong clustering property other than the one based on the OGP. Second,
and more importantly,
it is not entirely clear how to  link meaningfully the strong clustering property with m-OGP, which appears
necessary for many problems in order to bridge the computational gaps to the known algorithmically achievable
values. Likewise, it is not entirely clear what is the appropriate definition of either weak
or strong clustering
property in the ensemble version, when one considers a sequence of correlated instances, which is 
again another important ingredient in the implementation of the refutation type arguments. Perhaps some
form of dynamically evolving clustering picture is of relevance here. Some recent progress on the
former question regarding the geometry of the m-OGP was obtained recently in~\cite{ben2021shattering} 
in the context of the spherical spin glass model. We leave both questions as interesting open
venues for future investigation.

\section*{Discussion}
In this introductory article we have described a new approach to computational 
complexity arising in optimality studies of random structures. 
The approach, which is based on topological disconnectivity of the overlaps 
of near optimal solutions, called the Overlap Gap Property,
is both a theory and a method. As a theory it predicts  the onset of algorithmic 
hardness in random 
structures at the onset of the OGP. The approach has been verified at various 
levels of precision in many classes of problems
and the summary of the state of the art is presented in Table~\ref{table:OGP}. 
In this table   
the current list  of  models known to exhibit an apparent algorithmic hardness
is provided, along with references and notes indicating whether the OGP based 
method matches the state of the art algorithmic knowledge. 
The ''Yes'' indicates this to be the case, and ''Not known'' indicates that the  
formal analysis has not been completed yet.  
Remarkably, to the day
we are not aware of a model which does not exhibit the OGP, but which is known to 
exhibit some form of an apparent algorithmic hardness.

\begin{table}
\begin{center}
\scalebox{.7}{
\begin{tabular}{|l|c|c|}
\hline
Problem description & OGP matches known algorithms & References \\
\hline
Cliques in \ER graphs & Yes & Based on references below \\
Independent Sets in sparse \ER graphs & Yes & \cite{gamarnik2014limits},\cite{rahman2017local},
\cite{gamarnik2020lowFOCS},\cite{wein2020optimal},\cite{farhi2020quantumRandom} \\
Random K-SAT  & Yes  & \cite{gamarnik2017performance},\cite{coja2017walksat},\cite{bresler2021algorithmic}\\
Largest submatrix problem & Not known & \cite{gamarnik2018finding} \\
Matching in random hypergraphs & Not known & \cite{chen2019suboptimality}\\
Ground states of spin glasses   & Yes & 
\cite{gamarnik2019overlapAoP},\cite{gamarnik2020low}\cite{subag2018following},\cite{montanari2019optimization}, \cite{alaoui2020optimization},\\
Number partitioning & Yes (up to sub-exponential factors) & \cite{GamarnikKizildagNPP}\\
Hidden Clique problem & Yes & \cite{gamarnik2019landscape}\\
Sparse Linear Regression & Yes (up to a constant factor) & \cite{david2017high} \\
Principal submatrix recovery & Not known & \cite{gamarnik2019overlapGJS},\cite{arous2020free}\\ 
\hline
\end{tabular}
}
\end{center}
\caption{The up to date list of random structures known to exhibit the OGP and the apparent algorithmic harndess} 
\label{table:OGP}
\end{table}

The OGP based approach also provides a concrete  method for rigorously  establishing barriers for classes of algorithms. Typically
such barriers are established by verifying certain stability  (input insensitivity) of algorithms making them inherently
incapable of overcoming the gaps appearing in the overlap structures. While the general approach for ruling out 
such classes of algorithm
is more or less problem independent, the exact nature of such stability as well as the OGP itself is 
very much problem dependent,
and the required  mathematical analysis   varies from borderline trivial to extremely technical, often relying
on the state of the art development in the field of mathematical theory of spin glasses.

\section*{Acknowledgements} The support from the NSF grant DMS-2015517 is gratefully acknowledged.

\newcommand{\etalchar}[1]{$^{#1}$}
\providecommand{\bysame}{\leavevmode\hbox to3em{\hrulefill}\thinspace}
\providecommand{\MR}{\relax\ifhmode\unskip\space\fi MR }
\providecommand{\MRhref}[2]{%
  \href{http://www.ams.org/mathscinet-getitem?mr=#1}{#2}
}
\providecommand{\href}[2]{#2}

\bibliographystyle{amsalpha}

\end{document}